\documentclass[psfig]{aa}
\usepackage[dvips]{graphics}
\input psfig.sty

\begin{document}

\title{Inhomogeneous models for plerions: the surface brightness 
profile of the Crab Nebula}
\titlerunning{Inhomogeneous models for plerions: the surface brightness
profile of the Crab Nebula}

\author{E. Amato\inst{1,2}, M. Salvati\inst{3}, R. Bandiera\inst{3}, 
F. Pacini\inst{1,3}, L. Woltjer\inst{3,4}} 
\authorrunning{Amato et al.}

\institute{
Dipartimento di Astronomia e Scienza dello Spazio,
Universit\`a di Firenze, Largo E. Fermi 5, I--50125 Firenze, Italy
\and
Department of Astronomy, 547 Campbell Hall, University of California, 
Berkeley, CA 94720
\and
Osservatorio Astrofisico di Arcetri, Largo E. Fermi 5,
I--50125 Firenze, Italy
\and
Observatoire de Haute-Provence, F--04870 Saint Michel l'Observatoire,
France
}

\offprints{E. Amato,amato@arcetri.astro.it}

\date{}

\thesaurus{08(09.19.2, 09.09.1 Crab Nebula, 02.18.5)}

\maketitle

\begin{abstract}
We extend the
homogeneous model for the synchrotron emission from plerions to allow
more realistic predictions of the surface brightness distribution at 
different frequencies.
Abandoning the assumption of a uniform particle distribution, 
we assume that particles are injected into the synchrotron nebula
only in the vicinity of the pulsar. Their distribution is then
determined by MHD propagation in a magnetic field of spatially 
constant but time-dependent intensity.
This highly simplified model reproduces the integrated spectrum and
the synchrotron surface brightness profile of the Crab Nebula at 
radio frequencies. However, when applied to higher frequencies, it
underestimates the extension of the emitting region, suggesting 
that particle diffusion with respect to the 
magnetic field lines must occur.
\end{abstract}


\section{Introduction}

The integrated non-thermal spectrum of plerionic supernova remnants
is well accounted for within the framework 
of the homogeneous model by Pacini \& Salvati (1973, hereafter PS), 
who regard the plerion as an adiabatically expanding spherical bubble 
of magnetized relativistic fluid. The bubble is replenished by the 
continuous conversion of the pulsar's energy outflow into magnetic 
energy and relativistic particles. The energy balance of 
the plerion is determined by the competition between the decreasing 
pulsar's input and the losses the fluid undergoes because of radiation 
and expansion. Both particles and magnetic field are assumed to be
distributed uniformly through the bubble.

When the synchrotron emission of the Crab Nebula is calculated on the
basis of the PS model, both the decline of luminosity with time 
(V\'eron-Cetty \& Woltjer 1991) and the frequency of the low energy 
spectral break (Marsden et al.\ 1984) are consistent with
a magnetic field strength around $3 \times 10^{-4}$~G, in agreement
with estimates based on equipartition arguments (Woltjer 1958)
and also with recent measurements of the nebular inverse Compton
radiation flux (De Jager \& Harding 1992). Other applications of the
PS model include the interpretation of the overall spectral and 
evolutionary properties of other plerionic supernova remnants, as 3C58
and G11.2-0.3, which seem to be very different from the Crab Nebula 
(Woltjer et al.\ 1997).

Despite the success of the model at describing the integrated 
properties of plerions, the assumption of homogeneity is 
unsatisfactory. If as assumed the particles
are accelerated in the general environment of the pulsar and if loops of
magnetic field are created by its rotation, important gradients could be
expected in the particle and field distributions. Also observationally
such gradients are in evidence. The Crab Nebula does not have a uniform 
emissivity at any wavelength: it is concentrated towards the center, the 
more strongly so at the higher frequencies.

We shall here consider a modification of the PS model in which loops of
magnetic field are continuously injected with the relativistic electrons
attached to the field lines. Since little is known about the detailed
conditions around the pulsar we shall assume that the injection takes
place at a certain distance $r_0$ from the center. In the case of the Crab
Nebula the moving wisps do in fact suggest that the injection occurs at some 
$10''$ from the pulsar. Since the propagation velocity of magnetic 
disturbancies is likely to be much larger than the expansion velocity of the
bubble, we shall assume that the field adjusts itself to constant energy 
density, an assumption that is strictly justified only if the particle
pressure is below the magnetic pressure and the field is chaotic.

While in the case of the Crab Nebula the polarization studies at radio 
and optical wavelengths (Wilson 1972a, Woltjer 1958) indicate a 
predominance of azimuthal magnetic field, the modest degree of
polarization shows that there are different field directions along the
line of sight. Such a less regular field (and the effect of reconnection)
may well cause the particles to move more freely through parts of the 
nebula. Thus the present model, which includes only convection, is 
likely to be a limiting case. In fact we will find that the observed 
emissivity distribution is intermediate between the one computed here
and the homogeneous case.

\section{Outline of the model}
In our model the nebula is regarded as a sphere of radius $R(t)$,
expanding at a constant rate. A continuous
supply of new magnetic energy and particles is provided
by the central pulsar, whose energy output per unit time can be
written as
\begin{equation}
L(t) = {L_0 \over (1 + t / \tau)^\kappa}.
\end{equation}

Both new particles and new magnetic flux are injected into the nebula
within a spherical layer at a distance $r_0$ from the
central pulsar. We shall assume that the injection region has
zero extension, although very likely this is an oversimplification.

We assume that the particles' velocity distribution is isotropic and
that each particle emits essentially at its characteristic frequency:
$\nu_c =c_2 B_\perp E^2$, where $c_2=0.29 \times$
$(3 e c / 4 \pi (m c^2)^3)$ is a constant, 
$B_\perp = B {\rm sin} \theta$, and $E$ and $\theta$ are the particle's 
energy and pitch angle, respectively. Under these assumptions the nebular
synchrotron emissivity becomes
\begin{equation}
S_\nu(\nu,t,r) = {c_1 \over 2 c_2} \int_\Omega {d \Omega \over 4 \pi}
 \left( {B_\perp \nu \over c_2} \right)^{1 \over 2} N \left[ \left(
 {\nu \over c_2
 B_\perp} \right)^{1 \over 2}, t, r \right],
\end{equation}
with $c_1=(2 c / 3) ( e / m c^2)^4$ a constant and
$N(E,t,r)$ representing the particles' spectral and spatial density
at time $t$. The integral represents the average over particles' 
pitch angles.

As stated before, we regard the magnetic field as spatially constant,
therefore,
in our model, both the spatial and spectral characteristics of the 
nebular synchrotron emission are directly related to the particle 
spatial and spectral distribution, $N(E,t,r)$. 

$N(E,t,r)$ is related to the injected particles' spectrum through 
the particles' number conservation law:
\begin{equation}
N(E,t,r) = {J[E_i(E,t,r),t_i(E,t,r)] \over 4 \pi r^2} \left |{\partial
t_i \partial E_i \over \partial r \partial E}\right |,
\end{equation}
with $J$ being the number of newly injected particles per unit time
and energy interval and $E_i$ and $t_i$ being, respectively, the
initial energy and injection time of a given particle. The last
term in Eq.\ (3) represents the jacobian of the transformation
$(t_i,E_i) \rightarrow (r,E)$.

Apart from knowledge of the injected spectrum, in order to calculate
$N(E,t,r)$ from this equation, knowledge of the particles' age
($t-t_i$) and initial energy ($E_i$) as functions of their present
position ($r$) and energy ($E$) is needed.

We consider the energy evolution of each particle due to adiabatic and
synchrotron losses:
\begin{equation}
{dE \over dt} =
 -{E \over 3} \vec \nabla \cdot \vec u
 - c_1 {B^2}_\perp E^2.
\end{equation}
Once the magnetic field $\vec B$ and the fluid bulk velocity $\vec u$
are known at each time, Eq.\ (4) can be integrated to determine $E_i(E,r)$.

The determination of the time dependence of the magnetic field strength
$B$ is straightforward under our assumptions of time constancy of the
nebular expansion rate and spatial constancy of $B$ itself. Following
PS, we relate $B$ to the nebular content of magnetic energy
$W_B(t)$: $B^2(t)$=8 $\pi$ $W_B(t)$/ $V(t)$, with $V(t)=(4 \pi
/3)(R^3(t)-{r_0}^3)$ representing the confining volume for the
magnetic field. Then we write the time evolution of $W_B$ due to
expansion losses and the decreasing pulsar input:
\begin{equation}
{d W_B \over dt} = -{W_B \over 3} {d \over dt} \ln V(t) + \beta
 L(t).
\end{equation}
Integration of Eq.\ (5), where $\beta$ is a constant, yields the magnetic 
field strength at each time (see, for instance, PS) .

Knowing $B(t)$, and further assuming that the magnetic field is largely
azimuthal, $\vec B = B(t) \vec e_\phi$, we can use the flux freezing
condition,
\begin{equation}
{\partial \vec B \over \partial t} = \vec \nabla \wedge
\left( \vec u \wedge \vec B \right),
\end{equation}
as an equation for the fluid bulk velocity $\vec u$, which we assume
to be radial: $\vec u = u(r,t) \vec e_r$. Integrating Eq.\ (6) with the
boundary condition that the fluid velocity field matches the expansion
velocity of the nebula at its outer edge ($u(R(t),t)=v$), we find:
\begin{equation}
u(r,t)={v R(t) \over r}+{\partial {\rm ln} B \over \partial t}
{(R^2(t) - r^2) \over 2 r}.
\end{equation}
The velocity field starts at $r_0$ 
at a fraction of the speed of light of order $(v/c) \times (R/r_0)$ 
($\sim 1/10$, see below) and then it decreases roughly as 
$1/r$ to match the nebular expansion velocity at $R$.

Eq.\ (7) can also be written as
\begin{equation}
{d \over dt} [B(t)(R^2(t)-r^2(t))]=0,
\end{equation}
which yields, in its integral form:
\begin{equation}
B(t) \left( r^2- {r_0}^2 \right) =\\
\end{equation}
$$B(t) \left( R^2(t) -
{r_0}^2 \right) - B(t_i) \left( R^2(t_i) - {r_0}^2 \right).$$
This last equation simply states that the magnetic flux injected into
the nebula during the time interval $t-t_i$, with $t_i$ being the time
at which a particle that at time $t$ is at $r$ was born, is all
contained between $r_0$ and $r$. Taking $r=r(t,t_i)$ from Eq.\ (9), we
have got the relation needed to connect the particles' position at each
time to their age.

Presently what is left to determine, before being able to calculate
$N(E,t,r)$ as a function of the injected spectrum, is the evolution
of the particles' energy from initial to present value. Introducing
the expression found for $\vec u$ into Eq.\ (4), we obtain:
\begin{equation}
{d \over dt} \left[ {1 \over E} \left( {B(t) \over r} \right)^{1/3}
\right] = c_1 {B(t)^{7/3} \over r^{1/3}} {\rm sin}^2
\theta.
\end{equation}
We integrate Eq.\ (10) after replacing ${\rm sin}^2 \theta$ with
its time average. We take this to be 2/3, which is equivalent to 
state that each particle during its synchrotron lifetime experiences 
all the possible velocity orientations with respect to the magnetic 
field with the same probability. Obviously, we expect this 
approximation to work better the longer a particle lives, hence the 
smaller its initial energy is. Nevertheless we apply it to particles 
of all energies and finally find the expression for $E_i(E,t,r)$:
\begin{equation}
E_i (E,t,r) = E \left( {r \over r_0} \right) ^{1/3} \left(
{B(t_i) \over B(t)} \right)^{1/3} \left( 1-{E \over E_b(t,r)}
\right)^{-1}
\end{equation}
with
\begin{equation}
E_b(t,r)={3 \over 2 c_1} \left( {B(t) \over r} \right)^{1/3} \left(
\int^t_{t_i(r)} {B(t')^{7/3} \over r(t')^{1/3}} dt' \right)^{-1}.
\end{equation}
The meaning of the energy $E_b(t,r)$ is apparent: it represents
the maximum possible energy for particles that at time $t$ reside at
$r$. Substituting $E_b$ with $\sqrt {\nu / (c_2 B_\perp)}$, Eq.\ (11)
also defines the maximum radial distance $r_M(\nu,t)$ from which we 
expect emission to come for a given frequency. 

Inserting our findings into Eq.\ (3) we are finally able to relate
$N(E,t,r)$ to the injected particle spectrum. Concerning the
latter some assumptions are necessary. Assuming that the
injected electron spectrum is described by a single power law, 
synchrotron aging can account for just one break in the emission 
spectrum, while observation shows that in the case of the Crab Nebula
there is another break between the optical and the X-rays; this break
must be intrinsic to the injection mechanism.
Then we allow for a particle energy spectrum with 
two different slopes in two different energy ranges, namely:
\begin{equation}
J(E,t)= \left\{
 \matrix{ K_1(t)\ E^{-\gamma_1}\ ;\ \ E_m<E<E_1 \cr
 K_2(t)\ E^{-\gamma_2}\ ;\ \ E_1<E<E_M \cr
 0\ ;\ \ \ E<E_m\ {\rm or}\ E>E_M\ . \cr } \right.
\end{equation}

For simplicity, we assume all the energy cuts to be time independent
and the ratio between $K_1$ and $K_2$ to remain constant.
Since we have assumed that a constant fraction $\beta$ of the
pulsar's energy outflow goes into feeding the magnetic field, the 
remaining $1-\beta$ will be used for accelerating particles, and then
the constraint on the total slowing-down power converted into particles,
\begin{equation}
(1-\beta) L(t)=\int^{E_M}_{E_m} J(E,t) E dE,
\end{equation}
implies
\begin{equation}
K_1(t) \propto K_2(t) \propto L(t).
\end{equation}

Finally we have for the spectral and spatial distribution of particles
across the nebula:
$$N(E,t,r)= {\bar N \over (1 + t_i(r)/ \tau)^\kappa} E^{-\gamma} \left(
{r_0\ B(t) \over r\ B(t_i(r))} \right)^{\gamma-1 \over 3} \times $$
\begin{equation}
\left[ 1 -
{E \over E_b(t,r)} \right]^{\gamma-2} {1 \over 4 \pi r^2} {\partial
t_i \over \partial r},
\end{equation}
with 
\begin{equation}
\gamma = \left\{
\matrix{ 
\gamma_1 ,\ \ E_m(r)<E<E_1(r) \cr
\gamma_2 ,\ \ E_1(r)<E<E_M(r) \cr
}
\right .
\end{equation}
where the energies $E_m(r)$, $E_M(r)$ and $E_1(r)$ are the evolved
minimum, maximum and intrinsic break energy at radius $r$, respectively,
and can be calculated as functions of $E_m$, $E_M$ and $E_1$ by means of
Eq.\ (11).

\section{The case of the Crab Nebula}
For the Crab Nebula pulsar we have in Eq.\ (1) 
$L_0=3 \times 10^{39} \rm{erg/s}$, $\tau=710\ \rm {yr}$, 
$\kappa=(n+1)/(n-1)$ where the braking index $n$ is $n=2.5$ 
(Groth 1975).
For the parameters of the injected spectrum (Eq.\ (13)), following a
procedure similar to PS, we derive from the observations
the values $\gamma_1=1.54$, $\gamma_2=2.3$, 
$E_1 \simeq 20 \rm{eV}$. The parameter $\beta$ in Eq.\ (14) determines, 
through Eq.\ (5), the present value of $B$; we take the latter to be 
equal to $3 \times 10^{-4} {\rm G}$, so that $\beta=0.25$. 
Given these parameters the only unknowns in Eq.\ (16) are the injection
radius $r_0$ and $\bar N$, the latter containing the cut-off energies
of the injected spectrum. These parameters have very different 
observational signatures: the first one is related to the radial 
distance from the central pulsar of the luminosity peak, and
the second one only affects the overall nebular synchrotron flux.

The model predictions for the synchrotron surface brightness profile 
of the Crab Nebula have been compared with high resolution data at 
various frequencies.

For the radio band we have used a VLA map at a frequency of 1.4 GHz
(Bietenholz et al.\ 1997).

The spatial and spectral distribution of the optical synchrotron
continuum was determined by V\'eron-Cetty \& Woltjer (1993) after
subtraction of the thermal contributions from foreground stars and
filaments, from four narrow-band images at wavelenghts of 9241, 6450,
5364 and 3808 \AA. We have reanalysed these maps, kindly put at our
disposal by M.P. V\'eron-Cetty, with state-of-the-art star subtraction
algorithms.

Finally, in the X-ray band, we have used, after deconvolution of the
instrumental PSF and subtraction of the dust halo (Bandiera et al.\
1998), a collection of all the public ROSAT HRI data concerning the
Crab Nebula. We have estimated the mean photon energy of these data
to be 1 keV.

In our model
the synchrotron surface brightness $J_\nu$ is simply obtained by
integration of Eq.\ (2) along the line of sight:
\begin{equation}
J_{\nu}(\nu,t,z)= \int_z^{R(t)} S_{\nu}(\nu,t,r) {r dr \over
\sqrt{r^2-z^2}},
\end{equation}
with the expression for the particle number density $N$ in $S_\nu$
calculated from Eq.\ (16).

In order to compare our spherical model with the observations, we have
extracted from each image what we call a ``radial intensity profile'':
we first sampled the emission profiles of the nebula along different
directions, taking the mean values over small areas of $10'' \times 10''$
and then, after rescaling the different axes to a common length, we
averaged those profiles. The procedure just described, to which we refer
in the following as ``data sphericization'', is the main cause of
uncertainty in our radial profiles and it is what we take into account
in the error bars attached to the data points in most of the following
plots.

When the radiative losses are negligible ($E/E_b(r,t) \ll 1$ in Eq.\ (16)
throughout the entire nebula) $\bar N$ and $B$ enter the 
expression of the synchrotron emissivity simply as multiplying factors.
Therefore fitting the shape of the surface brightness profile at 
radio frequencies allows a straightforward determination of $r_0$, 
and once this is known, the value of $\bar N$ simply comes from 
fitting the integrated flux, so that the model is fully determined.

Our best fit estimate gave $r_0=10''$, which,
for a 2 kpc distance to the Crab Nebula, translates into $r_0=0.1$ pc.
The corresponding radial profile at 1.4 GHz is plotted against the 
data in Fig.\ 1. This value of $r_0$ yields a distance
from the pulsar to the injection site fully compatible with the 
association between the particle acceleration region and the location
of the optical wisps.

\begin{figure}[h!!!!!]
\centerline{\resizebox{\hsize}{!}
{\includegraphics{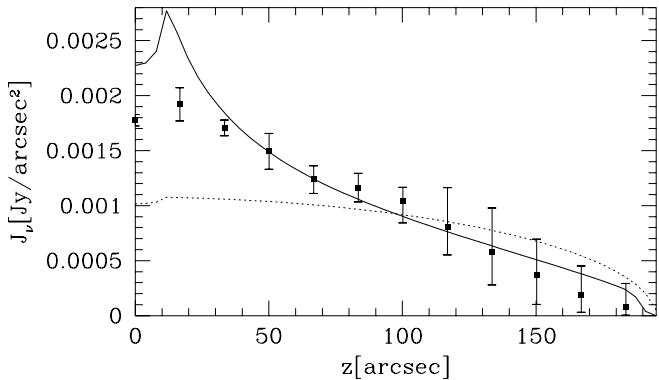}}}
\caption{\footnotesize{Comparison with the radio data (points) of the
PS model (dotted curve) and of our best fit model (solid curve) for 
the surface brightness profile of the Crab Nebula. The abscissa is 
the projected distance from the pulsar. The error bars attached to 
the data points take into account the uncertainties introduced by 
``data sphericization''.}}
\end{figure}

As in the PS model, the integrated fluxes reproduce the observations
at all frequencies: both the solid and the dashed curve yield the 
same flux as the interpolation of the data, when integrated on a 
spherical surface of radius 2~pc. Nevertheless, although in the radio
part of the spectrum the fit to the observed profile is rather good
and substantially improves the homogeneous model, the model 
predictions fail to reproduce the data at optical and X-ray 
wavelengths.

\begin{figure}[h!!!!!]
\centerline{\resizebox{\hsize}{!}
{\includegraphics{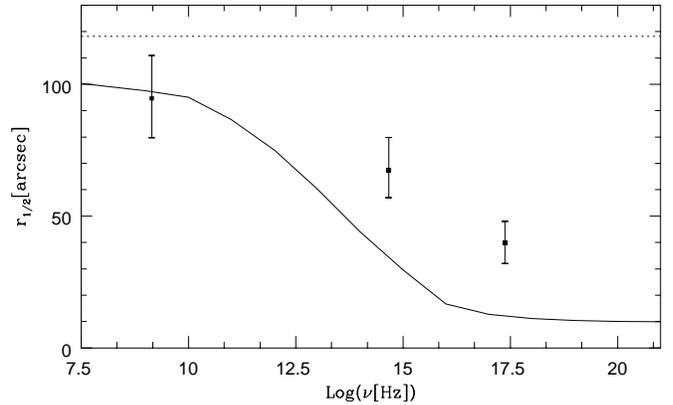}}}
\caption{\footnotesize{Comparison with the data at radio (1.4 GHz), 
optical (6450 \AA) and X-ray (1 keV photons) frequencies of the 
expected size of the emitting region. We plot as a function of 
frequency the radius within which half of the total nebular flux
per unit frequency is produced. The solid curve is based on the 
present model whereas the dotted curve represents the homogeneous
 PS model. Again the error bars attached to the data points take 
into account the effect of our ``sphericization'' procedure.}} 
\end{figure}

As shown in Fig.\ 2 the emission at optical and X-ray wavelengths
calculated on the basis of the present model is too concentrated: the 
highest energy particles emit most of their energy immediately after 
the injection. This causes most of the flux to originate from a narrow
region and the particles to travel a very short distance from the 
injection site before their energy is degraded by severe synchrotron 
losses.

This effect could be cured
by substantially lowering the magnetic field and thereby the synchrotron
losses. However the inverse Compton data do not allow this. Moreover if
the magnetic field energy is less than the particle energy the model
becomes invalid and at lower fields the total particle energy would
soon exceed that produced by the pulsar.

The homogeneous PS model in which the particles move freely through 
the Nebula yields too broad a distribution of the emissivity, our
model a too narrow one at the higher frequencies. Apparently some
of the particles can reach the outer parts of the nebula without 
suffering the large synchrotron losses which occur when they are
fully tied to the field lines.

Diffusion of particles could solve this problem, but the diffusion 
coefficient would have to be $\sim 10^{4}$ times larger than for Bohm
diffusion, in agreement with previous estimates (Wilson 1972b). 
A more complex field structure in which some lines connect the 
inner and outer parts of the nebula might give an acceptable solution,
since the particles could move along these lines at a substantial
fraction of the speed of light.

\acknowledgements{This work was partly supported by a grant from ASI.
We are indebted to Dr. M.P. V\'eron-Cetty  and Dr. M. F. Bietenholz
for putting at our disposal the optical and radio images of the Crab
Nebula, respectively. One of us (M. S.) acknowledges the hospitality of
the Department of Astronomy of the University of California at 
Berkeley, where part of this work was done.}

\end{document}